# Analysis of the Operation of Industrial Trucks based on Position Data

## Analyse des Betriebs intralogistischer Fahrzeuge basierend auf Lokalisierungsdaten

*Jakob Schyga*
*Hendrik Rose*
*Johannes Hinckeldeyn*
*Jochen Kreutzfeldt*

*Institute for Technical Logistics*
*Hamburg University of Technology*

**Indoor positioning systems (IPSs) can make an important contribution to the analysis and optimization of internal transport processes. The overall aim of this work is to examine how position data can be used to analyze the operation of industrial trucks in warehouses. This is achieved by developing a concept for the analysis of industrial truck operations based merely on position data. The concept consists of a signal processing scheme to derive kinematic data and three analysis methods – Monitoring, Area analysis, and Motion analysis. Schemes for the signal processing and detection of motion events were developed and implemented as part of the TrOpLocer-App (Truck Operation Localization Analyzer-Application) for recording, displaying, and processing position data, according to the proposed system concept. The TrOpLocer-App source code is published on GitLab [RS21]. Different filter algorithms were examined, as part of the signal processing scheme, from which the low pass Butterworth filter has shown the best results in static experiments. Validation of the motion detection scheme shows good detection quality for distinct events in a realistic movement experiment.**

*[Keywords: Analysis Concept, Indoor-Localization, Warehouse, Industrial Truck, Movement Detection]*

**Indoor-Lokalisierungssysteme (IPSs) können einen wichtigen Beitrag zur Analyse und Optimierung von innerbetrieblichen Transportprozessen leisten. Das übergeordnete Ziel dieser Arbeit besteht darin, zu untersuchen, wie Lokalisierungsdaten zur Analyse des Betriebs von Flurförderzeugen in Warenlagern genutzt werden können. Dies wird durch die Entwicklung eines Systemkonzeptes zur Analyse des Betriebs von Flurförderzeugen erreicht, welches ausschließlich die Verfügbarkeit von Positionsdaten voraussetzt. Das Konzept besteht aus einem Signalverarbeitungsschema zur Ableitung kinematischer Daten und drei Analysemethoden – *Monitoring*, Flächenanalyse und Bewegungsanalyse. Schemen zur Signalverarbeitung und zur Erkennung von Bewegungsereignissen wurden entwickelt und als Teil der TrOpLocer-App (Truck Operation Localization Analyzer-Application) zur Erfassung, Darstellung und Verarbeitung von Lokalisierungsdaten gemäß dem vorgestellten Systemkonzept implementiert. Der Quellcode der TrOpLocer-App ist auf GitLab veröffentlicht [RS21]. Es wurden verschiedene Filteralgorithmen als Teil des Signalverarbeitungsschemas untersucht, von denen ein Tiefpass-Butterworth-Filter in statischen Experimenten die besten Ergebnisse gezeigt hat. Eine Validierung des Bewegungserkennungsschemas zeigt eine gute Detektionsqualität für unterschiedliche Ereignisse in einem praxisnahen Bewegungsexperiment.**

*[Schlüsselwörter: Analysekonzept, Indoor-Lokalisierung, Warenlager, Flurförderzeug, Bewegungserkennung]*

## 1 INTRODUCTION

Efficient warehouses play a key role when it comes to maintaining the competitiveness of production and logistics companies and satisfying customer needs [JSS19]. Considering the increasing demands on delivery times, quality, and flexibility within the supply chain, the efficiency of warehouses is more important than ever. According to Richards [Ri14], a share of 22 % of a company's logistics costs can be traced back to warehouse operations. Thus, it is more important than ever to optimize the structure and all occurring transport processes in a way that safe and efficient operation can be guaranteed.

An essential component of warehouse operation is transportation. According to Burinskiene [Bu15], in 75 % of warehouses, goods are transported manually most of the time. Due to their ability to flexibly transport pallets and heavy goods, industrial trucks, such as forklift trucks and tugger trains are essential in means of transport [Al16]. Consequently, efforts are made to record all activities of industrial trucks to conclude the process quality and increase efficiency and safety in warehouses [Al16].

Indoor positioning is a technology for recording position data of people, vehicles, and other objects within warehouses. In addition to Warehouse Management Systems (WMSs) and Truck Control Systems (TCSs), Indoor Positioning Systems (IPSs) are a key data source for mining process information. Regarding the increasing digitalization, networking of value chains, and the development of the Internet of Things, IPSs can be used for a variety of logistic applications, such as navigation in robotics [Ya17], automated pallet booking [Ho14], or assets tracking [Zh17].

Literature research reveals the potential to increase process efficiency and safety within warehouses, based on the above-mentioned data sources, IPS, WMS, TCS, and additional sensors [Ha20], [Al16], [Al17], [HF19]. However, little attention is paid to the analysis of industrial truck operations, which is based merely on position data. The restriction in the use of position data results in advantages regarding integrability and maintainability. Heterogeneous fleets can be equipped temporarily or retrofitted with an IPS regardless of the vehicle, the WMS or the TCS used, which allows for simple, cost-efficient, and scalable solutions.

The overall target of the presented work is to examine how position data can be used to analyze the operation of industrial trucks in warehouses. Following this target, the following research question is posed (**RQ1**):

**How should a system concept look like, that allows the operation of industrial trucks to be analyzed merely based on position data?**

To obtain additional semantic information about the operation of industrial trucks and enable additional analysis methods, a motion analysis scheme based on position data must be integrated. Therefore, the following research question was additionally derived (**RQ2**):

**How should an analysis scheme, for the detection of motion events of industrial trucks merely based on position data look like?**

The publication is structured as follows: In literature existing approaches for the analysis of industrial truck operations are presented and categorized in Section 2. The investigation of the first research question (**RQ1**) is based on a concept design for analyzing the operation of industrial trucks, which is presented in Section 3. In Section 4 an answer to the second research question (**RQ2**) is given by presenting a scheme for signal processing and motion detection. The system concept including the developed methods and schemes is implemented into a software application, *named TrOpLocer-App* (**Tr**uck **Op**eration **Loc**alization Analyz**er**) being presented in Section 5. The developed signal processing and motion detection schemes are validated in Section 6. Finally, conclusions are drawn and an outlook on open research topics is given.

## 2 Analysis and Monitoring of Industrial Truck Operation

In the context of a warehouse, there are a large number of analysis and monitoring approaches that include different types of data, such as order data, movement data, position data, and vehicle status data. The data describing the processes of industrial trucks can be provided from global systems such as WMS, TCS, IPS, or sensors attached to the vehicles [Ha20], [Al17], [Al16]. The recorded data build the fundament for the determination of various Key Performance Indicators (KPIs) and the evaluation of the operational processes [Al16]. Figure 1 summarizes the analysis approaches for a possible industrial truck operation analysis application by categorizing different methods and KPIs according to the input data type and the operational goal. In the following, the categorized literature is briefly presented.

Alias et al. [Al16] developed a sensor package, which measures the forklift truck's activity and transmits it to a TCS. The velocity of the forklift truck is determined with a pitot-static system, which measures the air pressure while driving. Vertical acceleration sensors detect vibrations caused by the operation, from which the operating times are determined. An ultrasonic sensor detects the loading status of the fork. Based on the recorded data, different transport KPIs, such as the number and duration of operation and standstill times as well as the vertical movement of the fork and its usage are monitored.

Another approach to the analysis of industrial truck operation is provided by Al-Shaebi et al. [Al17]. Al-Shaebi et al. examine the influence of individual driving behavior concerning the productivity, safety, and energy consumption of the forklift truck. Various events are detected by measuring the velocity, acceleration, and battery status as well as the lifting velocity of the forks of a forklift truck. These are divided into acceleration and braking events, events of constant driving velocity, and events in which the fork was lifted or lowered. Based on the number and duration of driving events, the safety behavior of a forklift truck driver and the influence of driving behavior on the energy status of the forklift truck can be evaluated. As no position data is applied, the analysis of the driving behavior occurs only position-independent.

A holistic, system-integrative approach for recording the activities for a fleet of forklift trucks is proposed by

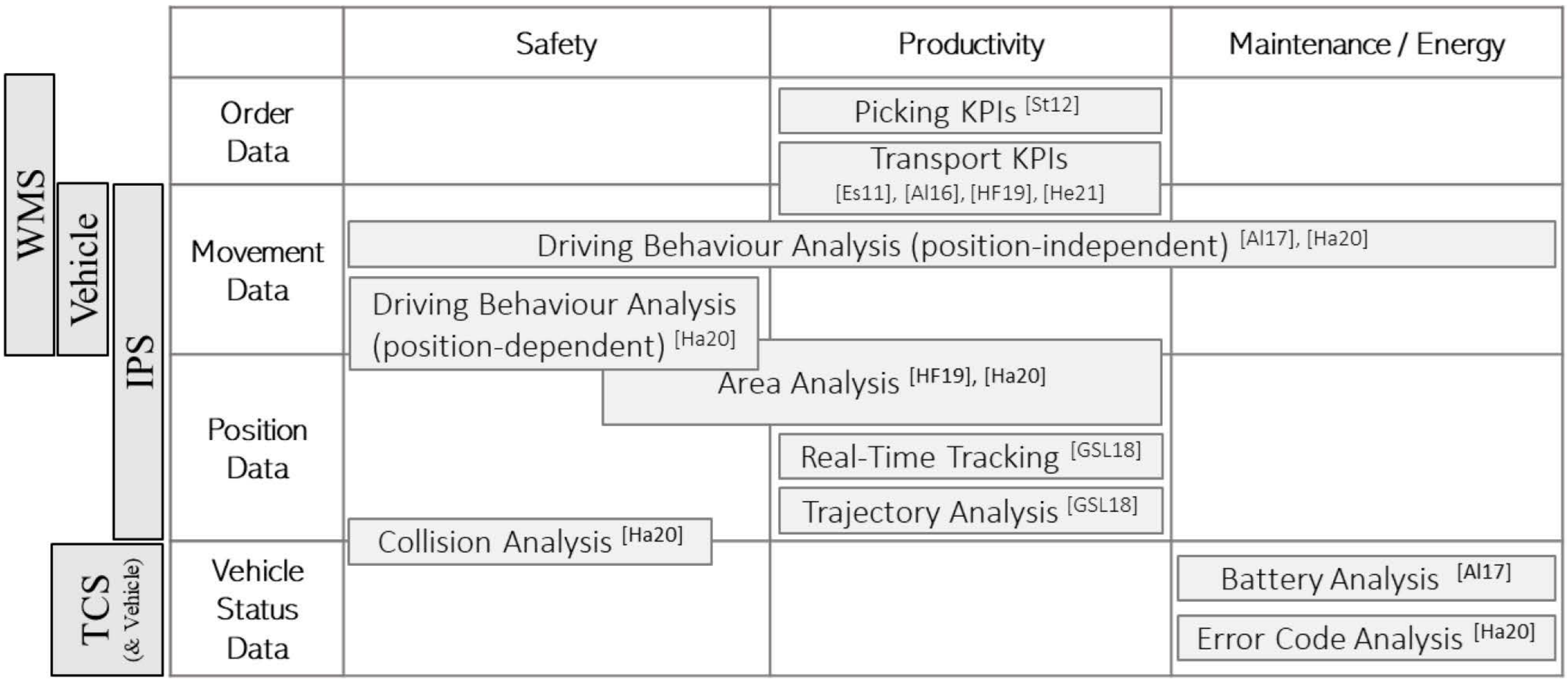

*Figure 1: Categorization of approaches for analyzing industrial trucks.*

Halawa et al. [Ha20]. Data is recorded from the WMS, the TCS, and an Ultrawideband (UWB) IPS. The WMS provides the database for order analysis, such as the number of orders, order processing times, and warehouse data. Forklift status data such as driver information and error codes from the internal forklift sensors are provided by the TCS. The position is recorded by the UWB system, which furthermore serves for the determination of the forklift trucks' velocity and acceleration. The integrative approach enables the combination of order-, movement- and vehicle-specific analyses. The evaluation of the warehouse processes takes place area- and vehicle-specific regarding the warehouse safety and operational efficiency including maintenance. The analysis includes brake harshness analyzes, congestion identification and prevention, compliance with routing policies and aisle strategies, analyzes of driving patterns at selected intersections, and the visualization of the location-specific distribution of the forklift stops and velocities. Error codes of the trucks are included to conduct impact and error analysis.

Hormes and Fottner [HF19] are providing another approach to combining the analysis of logistic vehicle processes and position data. The approach relates to the material flow analysis of a tugger train system by a UWB IPS. Travel routes, the warehouse layout, the cycle times of the orders, and the throughput of load carriers are analyzed. The analysis of the routes and the warehouse layout is supported by providing so-called heat maps, which assign the length of stay of the vehicles to individual sectors. Geofences are implemented to measure process times for the material flow analysis of the system.

So-called spaghetti charts show the trajectories of the tracked vehicles. They are widely applied for the analysis of production or logistics processes as presented in several case studies [HD18], [Da21]. Gladysz et al. [GSL18] present the generation of spaghetti charts for a forklift truck based on the data of a UWB IPS and the analysis of the operation in means of the simple differentiation of activities, real-time tracking, or reporting of the equipment utilization.

More advanced approaches for analyzing and further utilizing the data about industrial truck operation are presented by Ruppert and Abonyi [RA20], Estanjini et al. [Es11], and Hesslein et al. [He21]. Ruppert and Abonyi [RA20] present the integration of real-time position as well as acceleration data into digital twins for the simulation-based analysis of product-specific activity times. Estanjini et al. [Es11] demonstrate the optimization of forklift truck dispatching using sensors that collect vehicle information including location, usage time, accident history, and battery status. Transport KPIs in combination with the vehicle's position are also applied by Hesslein et al. [He21] to enable different location-based services, such as a position-dependent order placement service. Additionally, analysis methods can be based on the order and picking data. Approaches to describe manual picking efficiency are described by Stinson [St12].

As shown, various approaches, based on different data sources can be found in the literature, revealing the potential to gain transparency of industrial truck operation and therefore increasing warehouse safety and efficiency. However, a gap regarding the analysis of industrial truck operations merely based on position data and related research questions became apparent.

## 3 System Concept

This section aims at closing the identified research gap (**RQ1**) by designing a concept for the analysis of industrial truck operations based merely on position data. Instead of automatically generating concrete proposals for process improvements, the practical aim behind the concept is to gain transparency of the warehouse by providing information to process engineers for manual assessment, or information management systems as well as Automated Guided Vehicles (AGVs) or other cyber-physical-systems (CPSs) for automatic processing. The concept is explained below based on Figure 2.

Kinematic data such as velocity and acceleration can be determined from continuous position data, by calculating the first and second-order derivation of time. Measured position data is in practice scattering to a varying degree, which is amplified with each derivation, leading to extensive errors of the velocity and acceleration estimation. A signal processing chain, containing suitable filter algorithms must be provided to determine kinematic data while compensating for the scatter errors.

Considering position and kinematic data, different approaches from Figure 1 can be considered and supplemented by further functionalities. Analysis approaches from the literature, which are based on position or movement data are the *calculation of transport KPIs*, *position-dependent* and *position-independent driving behavior analysis*, *area analysis*, *trajectory analysis*, and *real-time tracking*. For the developed concept, the approaches are grouped into three independent analysis methods, which together build the analysis concept – *Monitoring, Area Analysis,* and *Motion Analysis*. The analysis methods are described in the following:

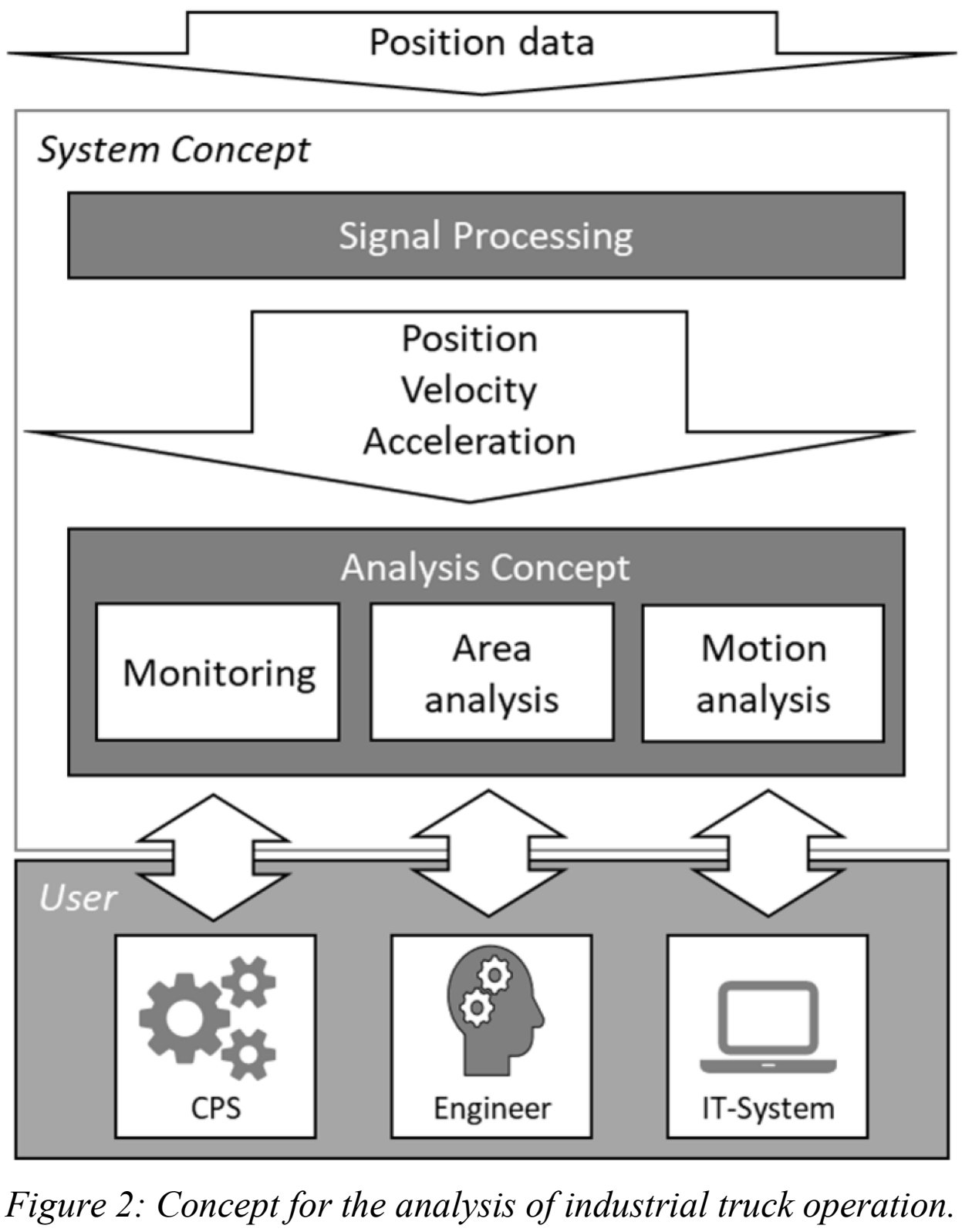

*Figure 2: Concept for the analysis of industrial truck operation.*

**Monitoring method**

The *Monitoring method* gives an overview of all tracked vehicles and their current positions on the warehouse map in real-time. This information can be applied manually e.g. to find a certain vehicle or to validate an assumed position. Furthermore, each vehicle's current velocity and acceleration can be monitored manually or further used by information systems or CPSs, e.g. to predict the arrival of goods or to inform an AGV about a possible danger of collision.

**Area analysis method**

The distributions of the vehicle's position, maximum velocity, and maximum acceleration are evaluated location-specific to generate intensity distributions, which can be visualized in heat maps. The intensity distributions are created using a two-dimensional grid with adjustable sector sizes. Furthermore, the *Area analysis method* holds the data to display the trajectories of the analysis objects. The information from the *Area analysis method* can be manually assessed or automatically processed, e.g. to analyze the utilization of an aisle, identify zones with heavy traffic, or safety-critical cross-sections.

**Motion analysis method**

The *Motion analysis method* includes all functions that relate to the movement of the analyzed objects including the detection of certain motion events and the determination of transport KPIs. The identification of the start time and the duration of the following motion events of industrial trucks are included in the concept, as they represent distinct events containing relevant process information:

- The *standstill* event is used to determine the downtime and service life of a vehicle within a specific observation period. The determination of *standstill* events and their duration in combination with the location can be used to analyze truck fleet utilization.
- A *maneuvering* event reflects all processes during which the vehicle was driving at low velocities for a longer period. The event depicts slow driving in aisles, movements during storage and retrieval from shelves, or the stacking of pallets. By recognizing *maneuvering* events in an aisle, for instance, the duration for storing and retrieving goods can be determined. Furthermore, *maneuvering* events on a travel route could indicate traffic congestions.
- *Driving* events exist to gain information about the travel durations, distances and locations of continuous travel. *Driving* is significantly different from *maneuvering* in the way, that no low velocities are reached and significant change in position occurs.
- *Harsh braking* and *strong acceleration* characterize dangerous driving behavior. Zones in which these

events accumulate are therefore to be assessed as safety-critical.

- Since forklift trucks can tip over when lifting heavy goods, and the driver's view can be restricted, *lifting or lowering the fork* is a safety-relevant process. If this event occurs simultaneously with driving, safety-critical driving behavior is indicated. This event can only be applied to forklift trucks.

Transport KPIs are calculated, based on the duration of the detected motion events. Selected transport KPIs are the *total driving time*, *the standstill time*, *the equipment utilization*, *the average driving velocity*, *simultaneous loading and driving,* and the *total driving distance*. The *total driving time* and the *total standstill time* result from the duration of the respective events, while the *equipment utilization* is determined from the ratio of standstill time to the total time considered. The *average driving velocity* is the arithmetic mean value of the velocities of all detected driving, braking, and acceleration events. The KPI *simultaneous loading and driving* provides information for a forklift truck about whether and how long the forklift truck has carried out lifting and lowering movements while driving.

The concept presented serves to answer the research question (**RQ1**) for a system design, that allows the operation of industrial trucks to be analyzed merely based on position data. The concept enables various analysis functions to be implemented in three analysis modules. The output from a system designed according to the proposed concept consists of various transport KPIs, kinematic data, intensity distributions, and trajectories segmented by motion events. The information can be manually assessed or further processed by information systems or CPSs. For the manual assessment by a process engineer, the data output could either be displayed numerically or in terms of colored spaghetti charts and heatmaps. The data queries and the configuration of the system are carried out by the users of the system.

## 4 Motion Analysis Scheme

The second research question (**RQ2**) of how an analysis scheme for the motion detection of industrial truck operation, based on position data should look like, is examined in the following. The motion analysis scheme is therefore divided into two sequential data processing schemes. The signal processing scheme enables the determination of kinematic data from position data and the motion event detection scheme subsequently serves to detect the given motion events.

### 4.1 Signal Processing Scheme

The signal processing serves for estimating the industrial trucks' kinematic data in real-time and for data log post processing. As previously discussed the position data has to be numerically differentiated twice. The following processing chain is proposed to be applied for the position data, resulting in velocity and acceleration values per timeframe.

1. Interpolation of position data
2. Application of data filter
3. Numerical differentiation
4. Application of data filter
5. Calculation of resulting velocity
6. Numerical differentiation
7. Application of data filter

The filter choice and parametrization play a key role, in the precise determination of the kinematic data. Three different filters are considered to enable the adaption of the signal processing according to the respective application and data quality. The Savitzky-Golay filter, an Infinite Impulse Response (IIR) Butterworth filter, and a Finite Impulse Response (FIR) filter [Py20], of which the IIR and FIR filter are possible candidates for real time data processing, were implemented. Those are initially suggested, as they represent common data filters, based on different operating principles. To compensate for the phase shift, the Butterworth and FIR filters have to be implemented as forward-backward filters, meaning that at each filter level, the data is again filtered in the opposite direction to achieve synchronization between position, velocity, and acceleration data for post analysis.

### 4.2 Motion Event Detection Scheme

Many approaches, ranging from simple clustering [Ch21], [SHZ21], [Fa21], to more advanced machine learning algorithms [Ha21], [SHZ21] for the segmentation of spatiotemporal data into events with semantic meaning, often referred to as trajectory mining, are widely discussed in the literature. In this section, a scheme for the detection of motion events based on the segmentation of time discrete position, velocity, and acceleration data by respective thresholds is presented. The scheme depicted by the block diagram in Figure 3 is described in the following.

According to the scheme, a sequence of different event calculation functions is applied to the given data input, consisting of the object's velocity (v), the resulting acceleration (a), and the vertical fork velocity ($v_{z,\,fork}$). Event calculation functions are applied for each of the event types defined in Section 3, in the order given in the table integrated into Figure 3. First, for each data frame in time, the necessary event condition is checked by comparing it with the chosen limit value. The default limit values are likewise given in the integrated table. After returning an event batch, the individual segments are corrected if necessary. On the one hand, a correction consists of the deletion of events, which fall short of a specified duration. On the other hand, the events can be checked for overlapping with already calculated events from the stack. If there is an overlap, the boundaries of the events are adjusted based on a set of rules. If all events are checked, the result is an event stack, which holds the detected events of each type. Finally, the data frames of each event stack are sorted according to time. The event *lifting/lowering of the fork* can occur simultaneously with the other events.

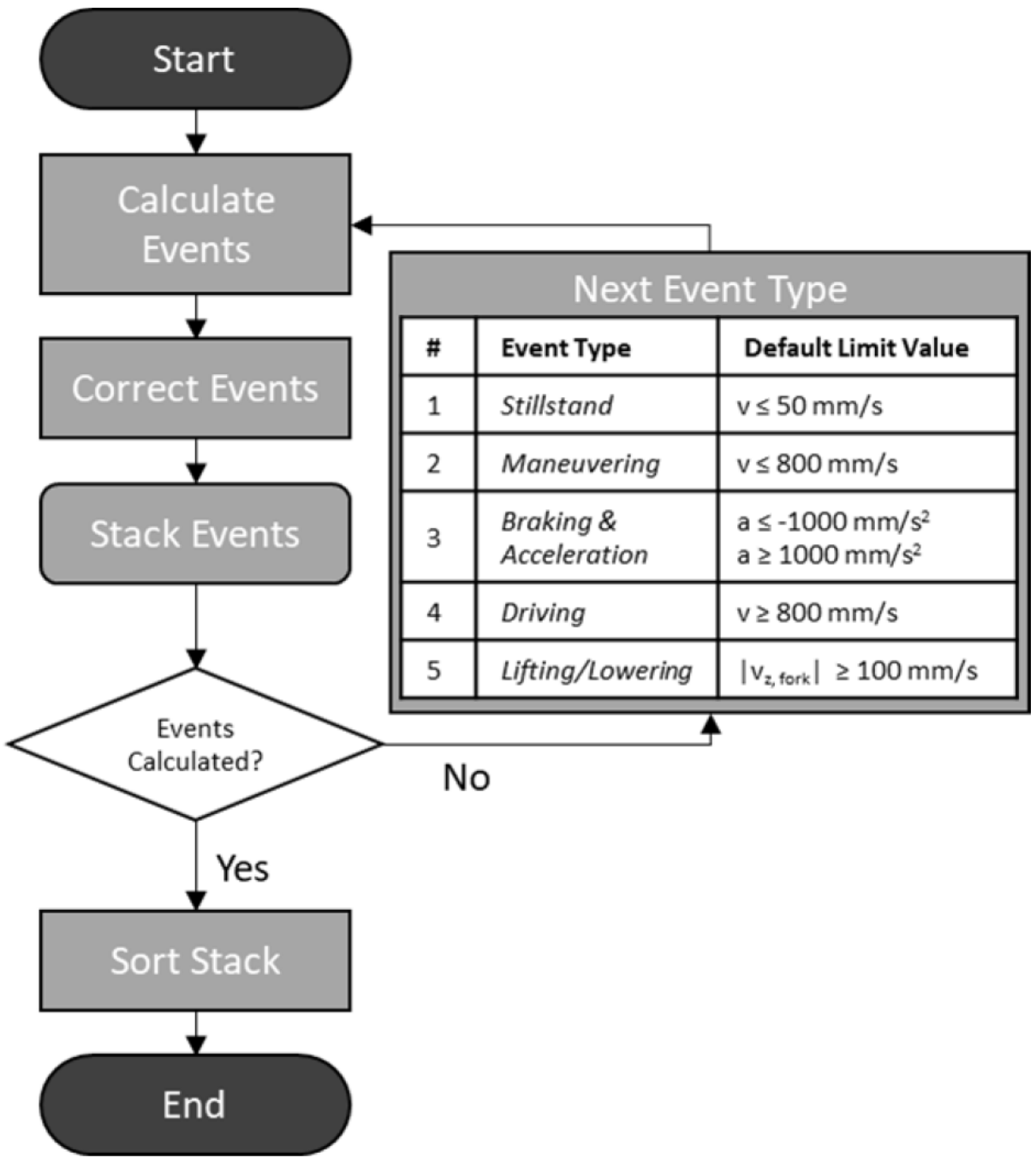

*Figure 3: Block diagram of the motion event detection scheme.*

In combination with the previously described signal processing scheme, the motion event detection serves to answer **RQ2**, by enabling the detection of the event types *standstill, maneuvering, harsh braking, strong acceleration, driving,* and *lifting/lowering of the fork* merely based on position data.

## 5 TrOpLocer-App

The presented system concept for the analysis of position data, including the presented schemes for signal processing and motion event detection, was implemented in the *TrOpLocer-App*. The *TrOpLocer-App* was developed particularly for the manual analysis of industrial truck operation by a process engineer but is designed in a way, in which information systems or CPSs could be linked in the future. The *TrOpLocer-App* source code is provided on GitLab [RS21].

Each of the analysis methods has been implemented as one of three modules, which are named according to the underlying method. The modules are connected and supplemented by several side components for reading in, configuring or exporting data, reading in and configuring floor plans, exporting visualizations, or configuring the analysis algorithms. The implementations of the Savitzky-Golay filter, Butterworth filter, and FIR filter from the Python SciPy library for signal processing [Py20] are integrated into the *TrOpLocer-App* application to enable the signal processing with adjustable filter parameters.

Figure 4 shows the main view frames of the graphical user interface. Besides the *start view*, each of the analysis methods of the prototype is implemented in a corresponding view frame.

In the *Monitor view* (B) the position (B1), velocity (B2), and acceleration (B3) over time are presented as a graph. Different buttons are implemented to control the data recording and processing. Numeric displays show the duration of the measurement, position and distance covered.

The *Area analysis view* (C) shows two main graphs. The left graph (C1) shows the position data of an object as a spaghetti chart. The right graph (C2) shows the evaluated position, velocity, or acceleration data as a heat map with different intensities for each sector of a grid. Both graphs can show different time segments of the evaluation by moving time sliders.

The *Motion analysis view* (D) shows an event map (D1) as the main component and a display for indicating the transport KPIs as numerical values (D2) on the right side. The underlying algorithm for the event detection is based on the presented motion detection scheme. The event map can display the selected events either as a colored trajectory or as a scatter distribution of the event starting position. A specific observation period can also be selected in the event map by moving the controls.

The *TrOpLocer-App* serves as an exemplary implementation for the manual analysis of industrial truck operation by integrating the concept and the schemes presented with visualizations and transport KPIs that allow for the manual analysis of industrial truck operation. Furthermore, the *TrOpLocer-App* is used as a tool in the following section for validating the presented analysis schemes.

## 6 Validation Experiments

This section aims to validate the presented analysis schemes by carrying out a static and dynamic experiment. The test area of the Institute for Technical Logistics at the Hamburg University of Technology serves as a test environment. The test area is equipped with a high-performance motion capturing system, covering a total area of about 100 $m^2$. With a recording frequency of 100 Hz, the system enables the real-time tracking of passive infrared markers with an absolute accuracy of less than 5 mm and scattering less than 2 mm [ADM17], [BFH16]. In motion capturing the angle measurements, taken by a minimum of two static infrared cameras with known positions, are used to determine the position of a reflective marker [Ra18]. A reach truck is used as an analysis object for experimental validation. The reach truck is equipped with passive markers on the roof.

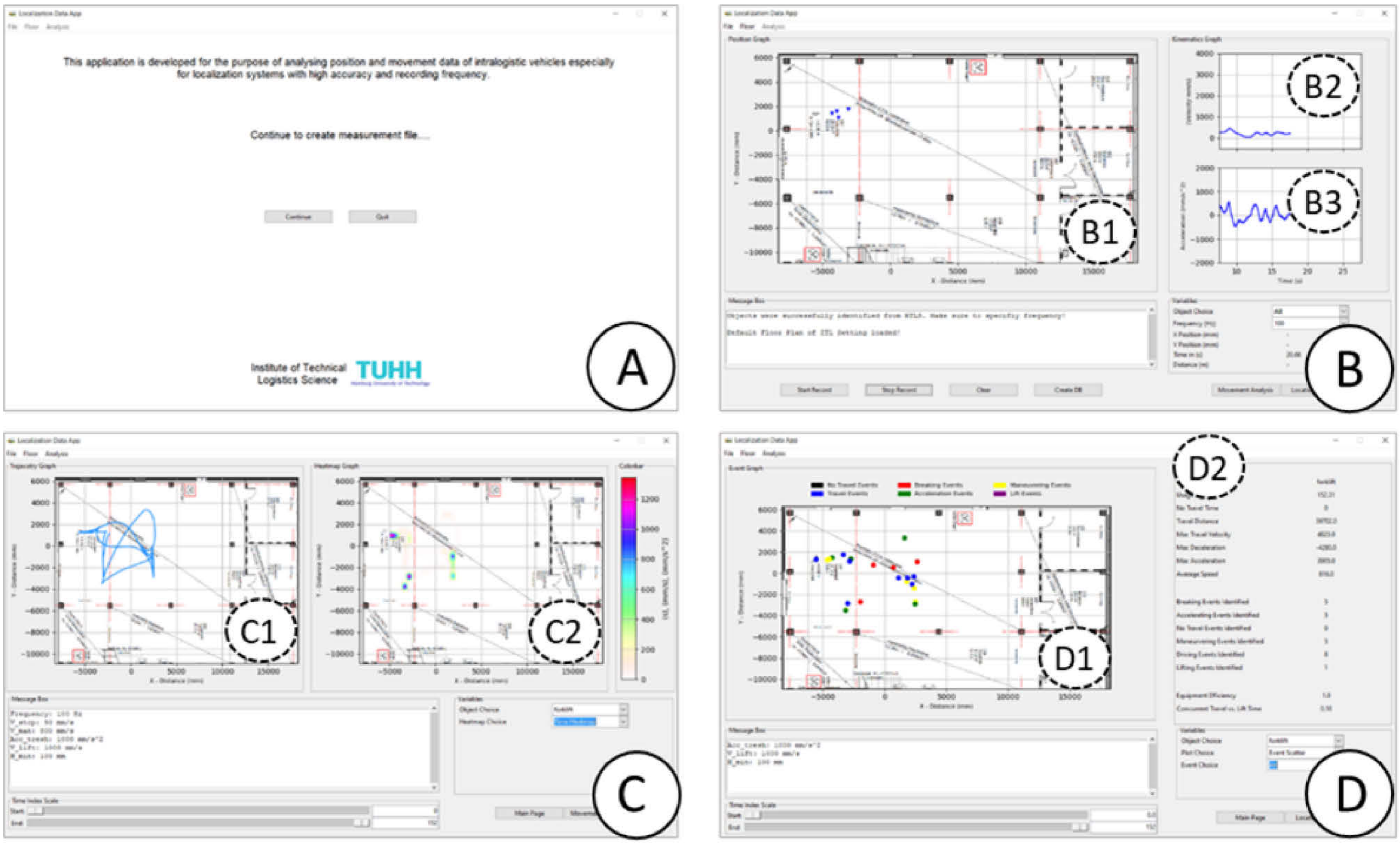

*Figure 4: Representation of the main views of the TrOpLocer-App: Start view (A, top left), Monitor view (B, top right), Area analysis view (C, bottom left), Motion analysis view (D, bottom right).*

### 6.1 Static Experiment

As all the implemented analysis methods from Figure 2 depend on kinematic data, the determination of the velocity and acceleration of industrial trucks based on position data is a key component of the developed concept.

By carrying out a static experiment, the reference velocity and acceleration are known to be zero. This is used for basic validation of the signal processing scheme with the different filter alternatives, integrated into *TrOpLocer-App*. If the output from the signal processing of a static experiment differs significantly from zero, the filter with the given configuration is not suitable for smoothing scattered data.

The experiment was carried out and the position data of the reach truck from the motion capture system was recorded by using *TrOpLocer-App*. The data is manipulated in terms of the update frequency and scattering of the position signal to examine the influence of the data input quality on the calculated velocity and acceleration. 1 Hz was chosen as the default limit frequency for the Butterworth and the FIR filter. For the recursive IIR filter a constant order of one was chosen. For the FIR and Savitzky-Golay filtering the length of the input signal (filter order) always covers a range of 0.5 s at the sampling frequencies to be examined. The default polynomial degree for the Savitzky-Golay filter is 2. Figure 5 shows the arithmetic mean values of the resulting velocity for each of the three filter types with varying update frequency and scattering. The scattering value describes the standard deviation of the artificial data manipulation, which is added to the motion capture position signal.

All filter alternatives show a good performance with an artificial degree of scattering of 0 mm. An average velocity of 800 mm/s is reached when using the FIR filter and the Savitzky-Golay filter. For the FIR filter, the resulting velocity begins to increase steadily from a scatter of 20 mm and reaches the maximum value of 800 mm/s with a scatter of 180 mm. The Savitzky-Golay filter shows a larger range for an average velocity of 800 mm/s, which extends over the frequency range from 5 Hz - 50 Hz. For the Butterworth filter, the defined maximum is not reached. The maximum velocity resulting from the Butterworth filter is 200 mm/s for a frequency of 5 Hz and a scatter value of 200 mm.

In conclusion, the FIR filter and the Savitzky-Golay filter with the given parameters require a higher quality of input position data regarding the update frequency and scattering and generate overall lower smoothing effects for the static position signals examined here. The IIR Butterworth filter is consequently suggested as the default filter for the described signal processing. The resulting inaccuracies, during a static experiment, have shown to be less than 200 mm/s with an update frequency of down to 5 Hz and an original scatter of up to 200 mm.

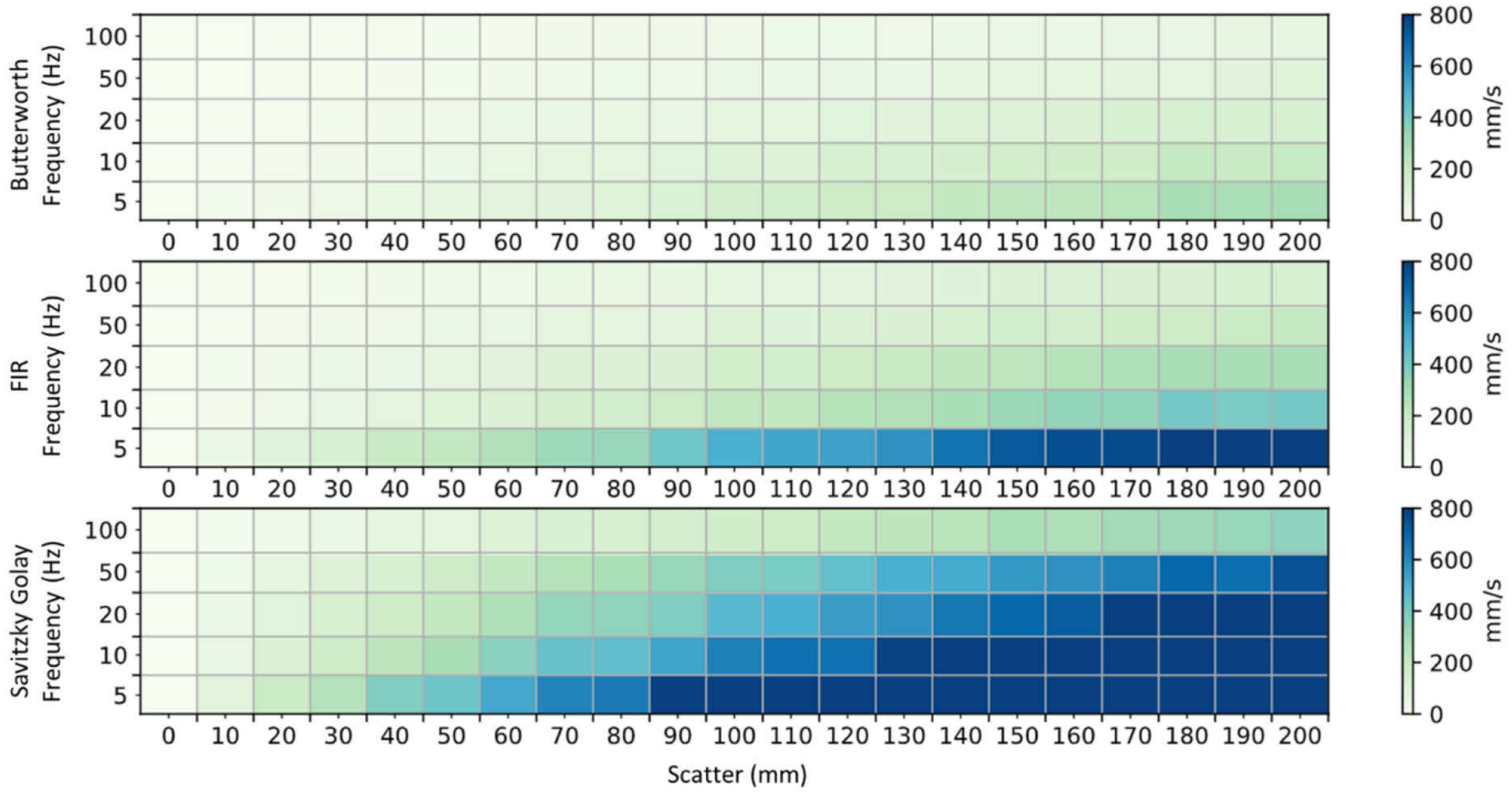

*Figure 5: Mean value of the resulting velocity of a static position signal after data manipulation.*

### 6.2 Movement Experiment

A movement experiment was designed and carried out to validate the motion detection scheme by applying the *TrOpLocer-App*. The test scenario includes specified forms of movement to be carried out within the test layout (Figure 6). The layout consists of different zones: A pallet shelf and shelving racks for carrying out storage and retrieval processes, a parking area for parking the vehicle and measuring downtimes as well as a loading and unloading zone for creating typical maneuvering situations. The remaining free space can be used for transports at different velocities. In addition, the diagonal, starting from the parking area, serves as the longest possible straight line for simulating harsh braking and strong acceleration as well as driving at high velocities.

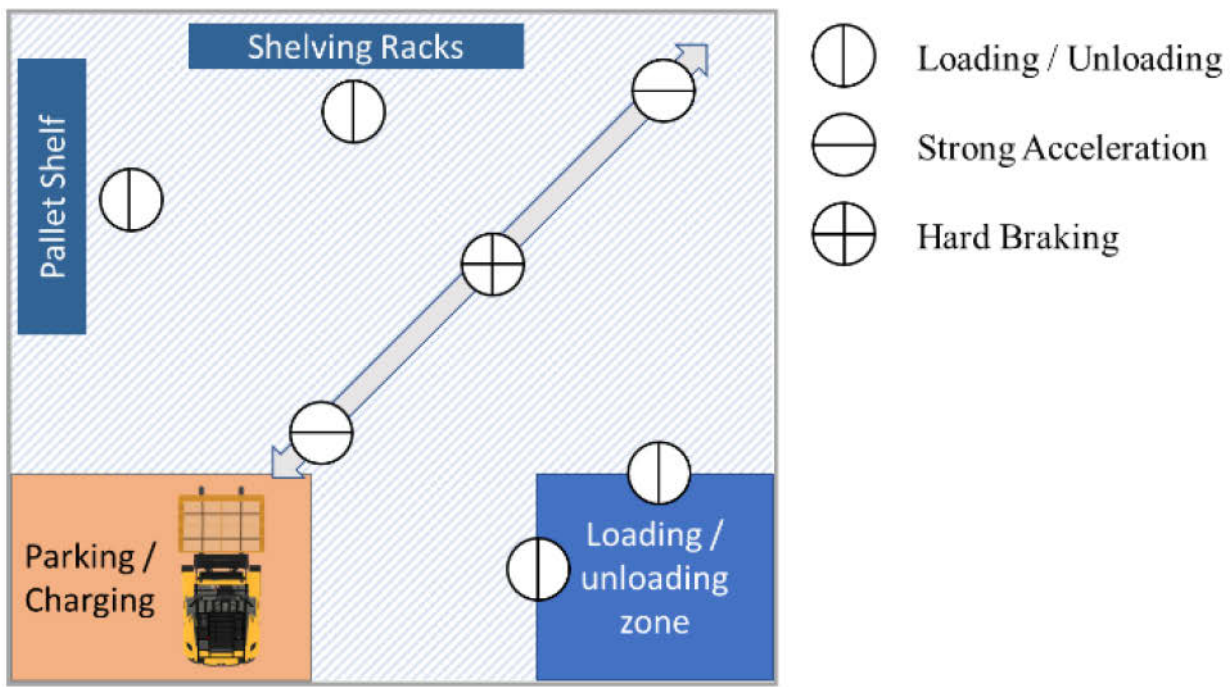

*Figure 6: Depiction of movement experiment.*

The position data of the movement experiment is recorded by the motion capture system and processed by applying the described signal processing from the *TrOpLocer-App* with the IIR filter. Subsequently, the presented motion detection scheme is applied based on the given default limit values.

As the assessment of motion events and transport KPIs is carried out by a process engineer, the algorithm's output should optimally resemble the personal perception of the processes. Following this argument, it is analyzed how accurate the detection of motion events by the *TrOpLocer-App*, matches the engineer's perception. The reference measurement consists therefore of several index sections that were manually created by two engineers based on a video analysis of the movement experiment. For each event type, when an event is identified, the video is paused to precisely determine the start and end index.

The segmented driving velocity curves from Figure 7 are applied for a qualitative analysis based on the comparison of the reference and the algorithmic event detection. The velocity curves are resulting from the signal processing algorithm of the measured position data of the motion capture system. The lines are colored according to the detected event section for the reference detection (above) and the detection algorithm (below).

All the *standstill, harsh braking,* and *strong acceleration* events of the reference are also covered by the detection. Further *standstill* events occur in the detection, which are identified in the reference as *maneuvering* events (Figure 7, A). Overall, there was a high level of agreement for the event type maneuvering. However, some event segments have been identified as *driving* events (Figure 7, B). In addition, there are differences concerning the transitions of individual successive section changes. The end indices of the maneuvering events at transitions to driving events are different when comparing the reference and detection results (Figure 7, C).

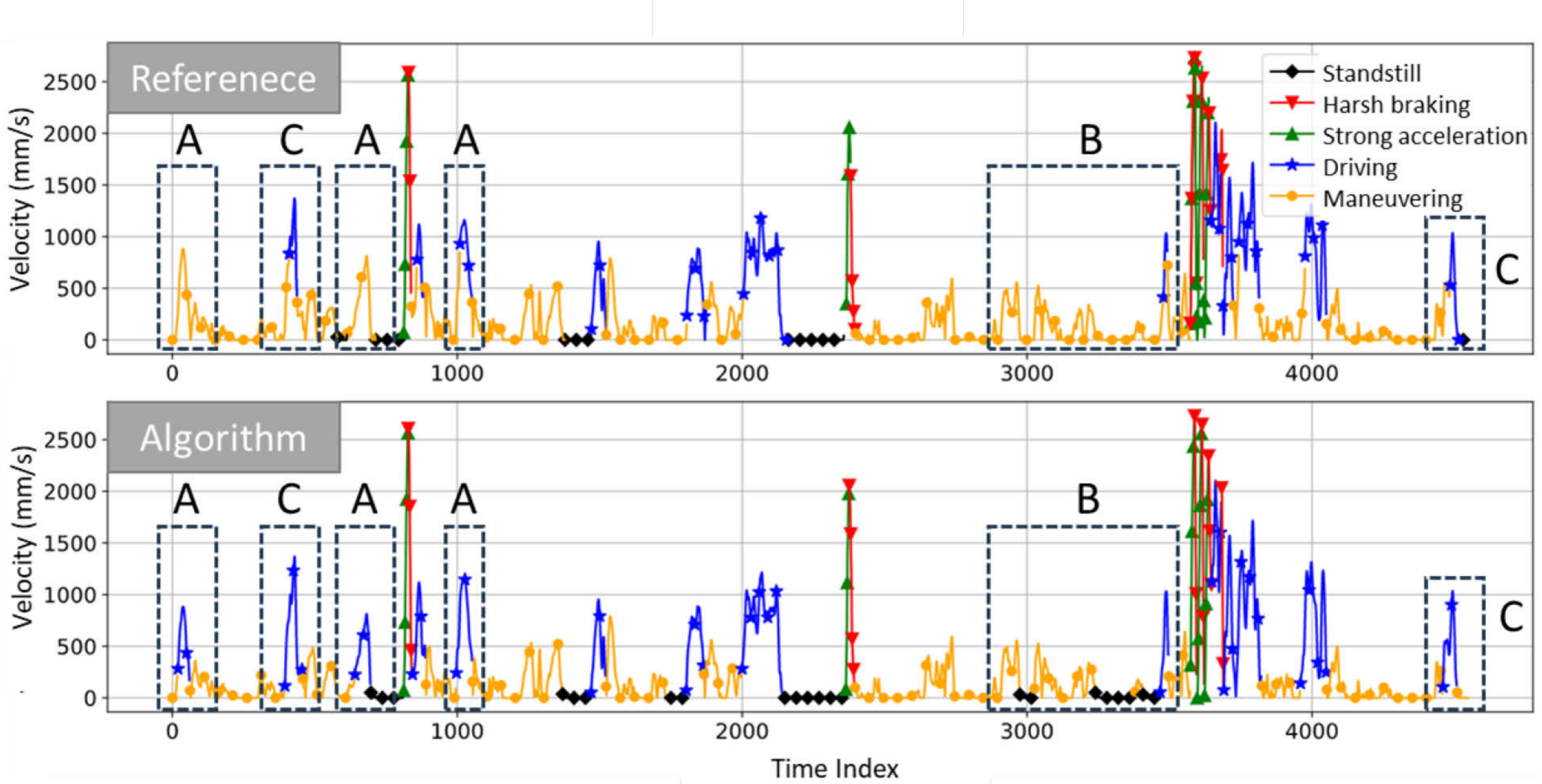

*Figure 7: Qualitative comparison of detected events of the reference (above) and detection algorithm (below).*

A generally valid statement about which movements correspond to a certain motion type cannot be made. However, it was shown that most motion events defined by the reference, were detected by the algorithm. Small differences between the start and the end indices can be neglected, as they are of minor practical relevance. While all *harsh braking* and *strong acceleration* events could be detected successfully, the differentiation between *driving*, *maneuvering,* and *standing still* is partly ambiguous. The neglection of *maneuvering events* would probably lead to a better detection ratio, but also significantly reduce the information content of the motion analysis.

The algorithm's detection performance could be further improved, by adapting the limit values accordingly. However, it was shown that the implementation of a simple rule-based motion detection scheme can successfully detect most of the motion events, based on high-quality position data.

## 7 Conclusions and Outlook

IPSs can make an important contribution to the analysis and optimization of internal transport processes, as they record the position of moving objects, such as forklift trucks, and thus transparently map the processes within a warehouse. The overall target of this work was to examine how position data can be used to analyze the operation of industrial trucks. The literature research has disclosed the existing approaches for analyzing industrial trucks, which are usually based on various data sources, such as WMSs, TCSs, IPSs, and vehicle sensors. However, limiting the data usage to position data reveals advantages, as the sole use of easy-to-install IPS also means that heterogeneous vehicle fleets can be easily equipped. Analysis can consequently be carried out independently of the vehicle's manufacturer or IT-infrastructure and temporary site surveys can become increasingly profitable.

A system concept was presented and therefore the research question (**RQ1**) of how to design a system, that allows the operation of industrial trucks to be analyzed merely based on position data was answered. The concept consists of a signal processing scheme for deriving kinematic data, and the three analysis methods – *Monitoring, Area analysis,* and *Motion analysis.* The schemes for signal processing and motion detection were subsequently developed to answer the research question of how a motion analysis scheme should look like, that processes merely position data (**RQ2**). The *TrOpLocer-App* is a software prototype, serving as an implementation of the developed system concept and the experimental validation of the developed schemes. The Butterworth, Savitzky-Golay, and FIR filters were examined, from which the Butterworth filter showed the best performance, regarding smoothening effects in the static experiment. A qualitative analysis of the implemented motion detection scheme revealed a high quality of detection for the motion events *harsh braking* and *strong acceleration* and a rather moderate quality for the events *driving, maneuvering,* and *standstill* when compared to human perception. The real-time position and kinematic data, the determined transport KPIs, the position, velocity, and acceleration distributions as well as the segmented trajectories and event locations can be manually assessed, e.g. to apply lean management methods [HD18], [Da21] or automatically processed to enable different location-based services, such as a position-dependent order placement service [He21] or for simulation as part of a digital twin [RA20].

The calculation of kinematic data is inevitable for the developed concept. The determination of the velocity and acceleration, based on position data with high scattering and low update frequency has proven to be faulty. The choice of an appropriate IPS for enabling the analysis functions is therefore not only limited by the position accuracy and real-time ability but furthermore by the system's update frequency and the data's scattering. Further evaluation criteria discussed in the literature are costs, complexity, scalability, reliability, or the availability on the market [Ya17], [MM16], [GLN09], [MPS15], [SAM17], [AK11], [HG12]. The motion capture system applied in the validations experiments does not fulfill the practical requirements for a suitable IPS, as the cameras are too cost-expensive and require continuous line-of-sight. Suitable systems, which are commonly applied in warehouse environments, could be camera-based systems [ÖAN16], [BHM18], UWB-systems [SAM17], [ZGL19], or systems based on Light Detection and Ranging (LiDAR) technology [We18], [BR14]. Accelerometers could be applied to directly determine the vehicles acceleration. However, the concrete system choice requires a more detailed investigation, e.g. by applying the *T&E 4Log framework* [Sc21], a framework for the application-driven test and evaluation of IPS in warehouses.

Overall, the presented system concept, the motion analysis scheme, as well as the *TrOpLocer-App*, offer various starting points for future research and development. A detailed comparison and investigation of filter alternatives with kinematic comparison data and additional filter options could provide information about the best possible filter option for generating kinematic data and thus optimize the signal processing. Furthermore, alternative schemes for motion detection could be examined or developed. In addition to the suggestions for improving the existing analysis, it is recommended that the approaches developed here are to be expanded into a holistic analysis, to further examine the potential in the use of position data. The literature research has shown that WMSs and TCSs provide useful information such as order information, throughput times, or error codes. Therefore, interfaces to these systems could be implemented in future work. Additionally, standardized interfaces e.g. to Omlox [Jö20] or RAIL [He21] as standardization initiatives for localization or the VDA 5050 [VD20] as a communication standard for AGVs could be integrated. To further analyze the potentials and limitations of the analysis of industrial truck operation based on position data, the application of the *TrOpLocer-App* will be further tested and evaluated in practical scenarios. At the time of writing, investigations are carried out, based on the UWB position data of tugger trains from a real warehouse environment (Figure 8).

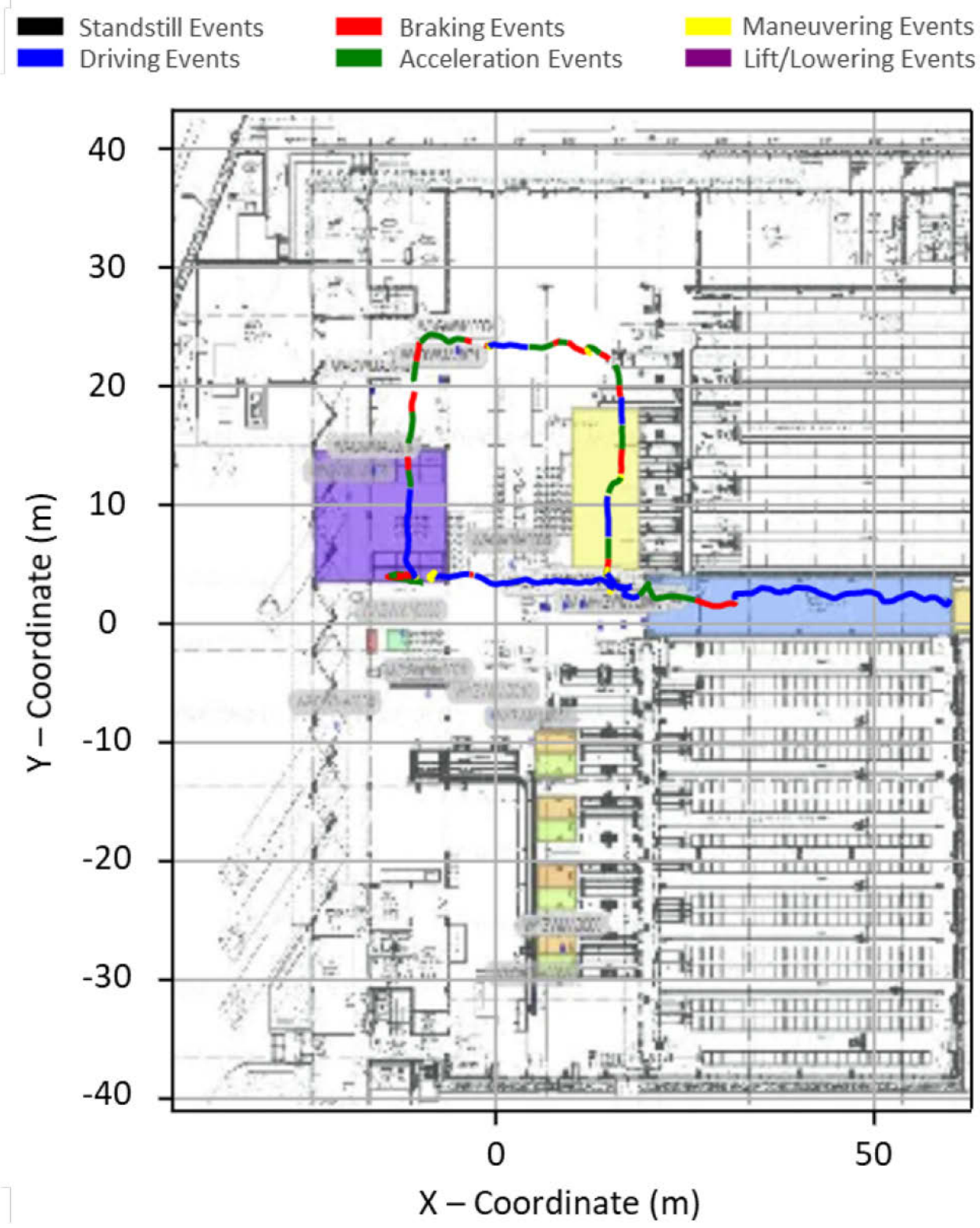

*Figure 8: Motion analysis view for UWB position data from a tugger train in a warehouse environment.*

---

**Jakob Schyga M. Sc.,** Research Assistant at the Institute for Technical Logistics, Hamburg University of Technology. Jakob Schyga studied mechanical engineering and production at the Hamburg University of Technology between 2012 and 2018.

**Hendrik Rose M. Sc.**, Research Assistant at the Institute for Technical Logistics, Hamburg University of Technology. Hendrik Rose studied general engineering and international industrial engineering at the Hamburg University of Technology between 2014 and 2020.

**Dr. Johannes Hinckeldeyn**, Senior engineer at the Institute for Technical Logistics, Hamburg University of Technology. After completing his doctorate in Great Britain, Johannes Hinckeldeyn worked as Chief Operating Officer for a manufacturer of measurement and laboratory technology for battery research. Johannes Hinckeldeyn studied industrial engineering, production technology, and management in Hamburg and Münster.

**Prof. Dr.-Ing. Jochen Kreutzfeldt**, Professor and Head of the Institute for Technical Logistics, Hamburg University of Technology. After studying mechanical engineering, Jochen Kreutzfeldt held various managerial positions at a company group specializing in automotive safety technology. Jochen Kreutzfeldt then took on a professorship for logistics at the Hamburg University of Applied Sciences and became head of the Institute for Product and Production Management

Address: Institute for Technical Logistics, Hamburg University of Technology Theodor-Yorck-Strasse 8, 21079 Hamburg, Germany; Phone: +49 40 42878-3557, E-Mail: jakob.schyga@tuhh.de